\def\kms{\ifmmode \mathrm{km~s}^{-1}\else km~s$^{-1}$\fi}
\def\msun{{\rm\,M_\odot}}
\def\point{   \parindent=20pt \par
              \hangindent\parindent }
\def\bullitem{   \point{$\bullet$\hskip2pt}}

\def\th{{\itshape th} }

\documentclass{iau-JDSS}
\usepackage{graphicx}

\title[Diffuse Light in Galaxy Clusters] 
{Diffuse Light in Galaxy Clusters }

\author[M. Arnaboldi and O.E. Gerhard]   
{Magda Arnaboldi$^1$ \and
Ortwin Gerhard$^2$}

\affiliation{$^1$ European Southern Observatory, Karl Schwarzschild-Str 2, 
85748 Garching, Germany \break email: marnabol@eso.org\\[\affilskip]
$^2$Max Planck Institute for Extraterr.\ Physics, Giessenbachstrasse,
85748 Garching, Germany \break email: gerhard@mpe.mpg.de}

\pubyear{2009}
\volume{Volume 15}  
\pagerange{119--126}
\date{?? and in revised form ??}
\jname{Highlights of Astronomy, Volume 15}
\editors{Ian F Corbett, ed.}
\begin{document}

\maketitle

\begin{abstract}
  Diffuse intracluster light (ICL) has now been observed in nearby and
  in intermediate redshift clusters. Individual intracluster stars
  have been detected in the Virgo and Coma clusters and the first
  color-magnitude diagram and velocity measurements have been
  obtained. Recent studies show that the ICL contains of the order of
  10\% and perhaps up to 30\% of the stellar mass in the cluster, but
  in the cores of some dense and rich clusters like Coma, the local
  ICL fraction can be high as 40\%-50\%. What can we learn from the
  ICL about the formation of galaxy clusters and the evolution of
  cluster galaxies? How and when did the ICL form? What is the
  connection to the central brightest cluster galaxy? Cosmological
  N-body and hydrodynamical simulations are beginning to make
  predictions for the kinematics and origin of the ICL. The ICL traces
  the evolution of baryonic substructures in dense environments and
  can thus be used to constrain some aspects of cosmological
  simulations that are most uncertain, such as the modeling of star
  formation and the mass distribution of the baryonic component in
  galaxies.
\keywords{(cosmology:) large-scale structure of universe;
          galaxies: clusters: general;
          galaxies: evolution;
          galaxies: interactions;
          galaxies: structure;
          galaxies: kinematics and dynamics; 
          galaxies: star clusters;
          (ISM:) planetary nebulae: general}
\end{abstract}

\firstsection 

\section{Introduction}

The Joint Discussion dedicated to the study of diffuse light in
clusters took place on the 6\th and 7\th of August, 2009 during the
{\itshape XXVIIth} IAU General Assembly in Rio de Janeiro. It was the
first scientific meeting on this subject. This Joint Discussion
provided a forum to confront observational evidence and theoretical
predictions, and to identify future directions for understanding the
origin and implications of this new component of galaxy clusters.

The Joint Discussion included four main sessions covering the
distribution of diffuse light in clusters and groups, the kinematics
of intracluster stars, the intracluster stellar populations, and the
predictions of cosmological simulations for the evolution of galaxies
in clusters and groups, and for the formation of the intracluster
light. 14 invited plus 10 oral talks, and 14 poster papers contributed
to an intense scientific exchange and set the stage for a lively
scientific discussion, which concluded the workshop.

In what follows, we provide a brief summary and some selected
references for the talks in the four sessions, and end with a
summary of the discussion which took place on August 7\th, 2009.

\section{Distribution of diffuse light in cluster and groups}\label{sec:DL}

\subsection{Clusters at $z=0$}\label{subsec:cluz=0}

\subsubsection{Diffuse light in the Virgo cluster - C. Mihos} 
Thanks to the specially adapted Burrel Schmidt telescope, the study of
intracluster light in the Virgo cluster reached for the first time a
photometric accuracy of significantly less than $1$\% of night sky
emission over an area of many square degrees.  Deep V ($\mu_V= 28.5$
mag/arcsec$^2$) and B band ($\mu_{B} = 29.0$ mag/arcsec$^2$)
photometry with calibrated, quantitative photometric solutions and a
well understood error model were obtained. This work has revealed
faint surface brightness features over a multitude of angular scales,
from narrow streams to extended diffuse halos.  The B-V colour of the
ICL is similar to that in the outer halos of ellipticals, except for
some streamers, and there is a rough correlation with the spatial
distribution of intracluster planetary nebulae. References: Mihos et
al.\ (2005, 2009).

\begin{figure}
\includegraphics[height=3in,width=4in]{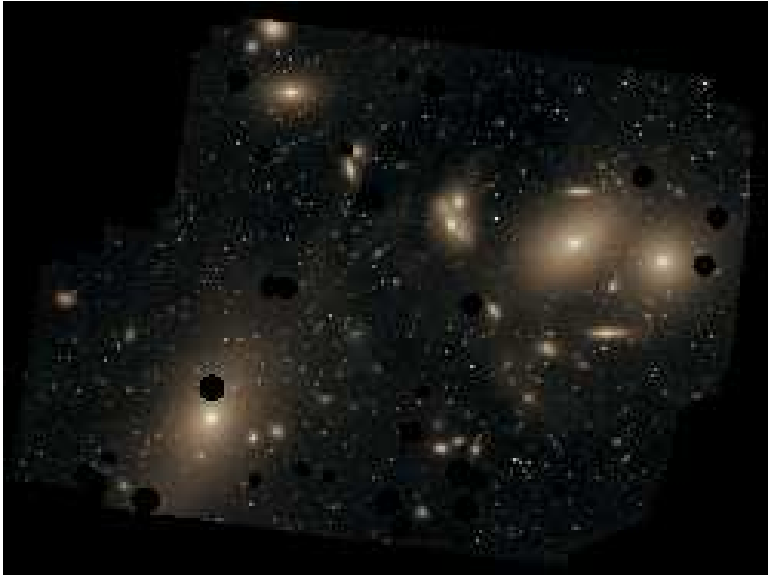}
  \caption{This deep image of the Virgo Cluster obtained by Chris
    Mihos and colleagues using the Burrell Schmidt telescope shows
    the diffuse light between the galaxies belonging to the
    cluster. North is up, east to the left. The dark spots indicate
    where bright foreground stars were removed from the image. 
    M87 is the largest galaxy in the picture (lower
    left). From ESO PR19/2009, based on Mihos et al.\ (2005).}
    \label{fig:mihos}
\end{figure}

\subsubsection{ Diffuse light in the Coma cluster - C. Adami} 
Diffuse light in the Coma cluster is documented since the work of
Zwicky in 1951.  Photographic photometry in the 1970s revealed a large
elongated diffuse component in the core of the cluster, around the two
BCG galaxies.  Several small-scale streamers and plumes were
discovered in the 1990s with CCD photometry.  The CFHT multi-color
Coma survey shows that the diffuse light is found either in the centre
of the Coma cluster or along the in-fall directions towards nearby
large scale structures.  The diffuse light in the cluster core was
found to be distributed on different scales, as quantified by a
wavelet analysis.  References: Thuan \& Kormendy (1977), Adami et al.\
(2005a, 2005b).

\subsubsection{Testing dark matter with the ``star pile'' in Abell
  545: the faintest cD or the brightest ICL? - R. Salinas} 
The study of surface brightness distribution of the star pile in Abell
545 shows that it can be described as the halo of a cD without the
high surface brightness central galaxy. The light distribution in
Abell 545 can thus be considered as independent evidence for
intracluster light as a separate stellar component from the BCG
halo. The spectroscopic follow-up shows that its spectrum is
consistent with that of an old stellar population. References: Salinas
et al.\ (2007).

\subsection{Groups at z=0: compact and fossil groups}\label{subsec:groupz=0}

\subsubsection{Diffuse light and intra-group star-forming regions in
  compact groups of galaxies - C. Mendes de Oliveira} Compact groups
are high density regions in the universe where the morphology of both
the stars and the HI gas in galaxies is often disturbed, providing
evidence for on-going interactions. An evolutionary sequence for
compact groups can be traced by the fraction of light in the diffuse
component on the group scale, the intragroup light (IGL), which
increases with the degree of interactions, reaching up to 30-40\% in
groups with many ongoing interactions.  Groups with the highest
fraction of IGL light are also those with the highest fraction of
early type galaxies. The colour of the IGL in most cases is similar to
the colours in the outskirts of the member galaxies.  Star formation
occurs in the intragroup medium when the compact group contains
stripped HI and is in an intermediate to advanced stage of
interaction.  References: Da Rocha et al.\ (2008), de Mello et al.\
(2008), Torres-Flores et al.\ (2009).

\subsubsection{Wavelet analysis of diffuse intra-group optical light in 
compact groups of galaxies - C. Da Rocha} 
The wavelet analysis of diffuse optical intra-group light in compact
groups of galaxies can provide reliable measurements of the intragroup
light on different scales. The diffuse light distributions in HCG 79
and HCG 51 are illustrated as prototype cases. References: Da Rocha et
al.\ (2008).

\subsection{Clusters at z=0.3}\label{subsec:clu=0.3}

\subsubsection{Intracluster light in moderate redshift clusters - 
J.J. Feldmeier} 
The review of photometric measurements of ICL fractions in clusters
highlights the challenges posed by such measurements and their
intrinsic limitations. The ICL is difficult to measure, being at best
at the level of 1\% of the night sky. Critical steps in the data
reduction are the sky subtraction and flat fielding, which have to be
precise to 0.1\%. Severe obstacles are the effects of the large scale
PSF and scattered light, and the separation of ICL from BCGs, cD halos
and other galaxies. Fractions of the ICL are determined from
photometric measurements, integrating the light outside the assumed
outer radii of individual galaxies over the region confined within an
isophote with surface brightness threshold defined by the depth of the
observations. Such measurements for the ICL fractions are in the range
from 10 to 45\% - ICL is common in galaxy clusters. References:
Feldmeier et al. (2004a), Gonzalez et al. (2007), Krick and Bernstein
(2007).

\subsubsection{Statistical properties of the intracluster light 
at z~0.25 - S. Zibetti}
To quantify the amount of ICL and determine trends with cluster
properties, key issues are statistics and depth. These can be
addressed by stacking many shallow images, combining them into deeper
ones. This procedure was adopted for a sample of 683 clusters with
BCGs from the SDSS sample. Average images were obtained after masking
all detectable sources (foreground stars and galaxies), centering and
scaling; they provide average properties of the diffuse optical light.
The results for the integrated light fractions within 500 kpc cluster
radius are BCG:ICL:galaxies = 21.9:10.9:67.2, with a robust estimate
of the fraction (ICL+BCG)/(total light) being about 30\%. The color of
ICL is similar to the color of galaxy light in the cluster, and the
radial profile of the ICL is more centrally concentrated than that of
the light in cluster galaxies.  References: Zibetti et al. (2005),
Pierini et al. (2008).

\subsubsection{ The diffuse intergalactic light in intermediate
  redshift clusters : RXJ0054-2823 - J. Melnick} An ``S''- shaped arc,
bluer than the cD galaxies and the ICL, has been detected in the
intermediate redshift cluster RXJ0054-2823 using the consecutive
differential image technique. There are no emission lines associated
with this arc and its redshift is consistent with the cluster
redshift, i.e. it is not a lensed background galaxy image.  The arc is
probably formed by two spirals caught in the act of being tidally
crunched by the three giant galaxies in the cluster
center. References: Melnick et al.  (1999), Faure et al. (2007).

\section{Kinematics of Intracluster stars -  individual stars and absorption
 line spectroscopy}\label{kin}

\subsection{Properties of intracluster starlight - A. Zabludoff}
The intracluster stars are hard to count: yet, they are significant in
the understanding of the origin of baryon vs.\ dark matter
distributions and the enrichiment of stars and gas in clusters.  ICL
is here detected in terms of an additional de Vaucouleurs profile
required to fit the surface brightness profile of the central galaxy
at large radii. A two-dimensional fit with a single profile is often
poor at large radii, and fails both for the ellipticity and position
angle profiles.  The BCG and the outer component are aligned within 10
deg about 40\% of the time; in the rest of the cases the misalignment
is large.  The extrapolated light of the outer component dominates the
total and its color is similar to that of an old population. The
increase of the velocity dispersion measured in several clusters
suggests that it responds to the cluster potential. References: Kelson
et al.\ (2002), Zaritsky et al.\ (2006), Gonzalez et al.\ (2007).

\subsection{Intracluster planetary nebulae and globular clusters}\label{sec:kinPNe} 

Discrete objects like planetary nebulae (PNe) are excellent tracers
for measuring the line-of-sight (LOS) kinematics of intracluster
stars: PNe occur during the final phase of solar type stars, their
number density distribution follows light, and their nebular shell
re-emits more than 10\% of the UV light from the stellar core in one
optical emission line, [OIII]5007\AA. This bright emission line makes
it possible to detect and measure the velocity of such stars even in
regions where the total surface brightness is too faint for absorption
line spectroscopy.  By surveying large areas in clusters, a suitable
number of PNe can be detected and mean radial velocities and velocity
dispersions can be determined.

Such investigations have been carried out in a number of nearby clusters
and this section covers the recent observational results in this
field. Similar studies are possible using globular clusters, assuming
these also trace the distribution of stars. Studies based on globular
clusters are reported in Section \ref{sec:pop}.

\subsubsection{Kinematics of diffuse light in the Virgo cluster core from planetary Nebulae - M. Arnaboldi}

\begin{figure}
\includegraphics[height=3in,width=3in]{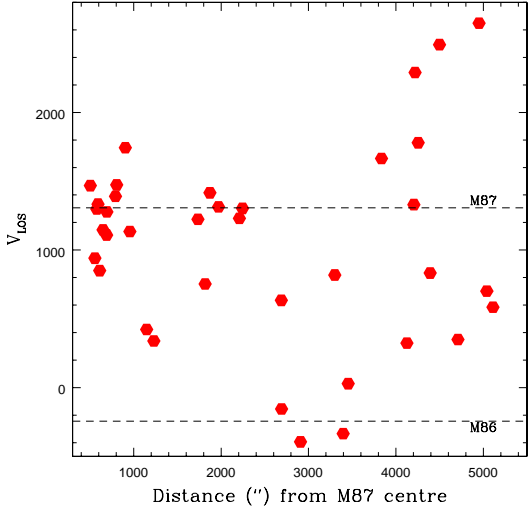}
  \caption{Distribution of line-of-sight velocity versus projected
    distance from the center of M87 for all spectroscopically
    confirmed PNs in the 6 FLAMES fields in the Virgo cluster
    core. From Doherty et al.\ (2009).}\label{fig:virgophsp}
\end{figure}

The kinematics of the ICL in the Virgo cluster was studied using the
PNe identified in narrow band surveys. The FLAMES spectrograph at the
VLT was used for the spectroscopic campaign, with a total of 6
VLT-Flames pointings observed.  From the positions and velocities of
the detected PNe a projected phase-space diagram was built which
illustrates the different dynamical components along the LOS in the
Virgo cluster core, see Fig.~\ref{fig:virgophsp}: these are the
extended halo of M87 and the ICL component. Within $R< 3600'' = 260$
kpc from M87, the ICL component is all at negative velocities relative
to M87, and it is not phase-mixed.  Part of this component is at
velocities consistent with the idea that light of the M86 sub-group is
being tidally stripped by the more massive M87 component, while the
two galaxies approach each other along the LOS. By contrast, in a more
distant field at $R=4500''$, the full velocity width of the Virgo
cluster is seen.  References: Arnaboldi et al.\ (1996, 2004), Doherty
et al.\ (2009).

\subsubsection{Dynamics of cluster cores and brightest galaxies from
  planetary nebulae velocities - O. Gerhard}
Numerical simulations predict that isolated galaxies acquire extended
radially aniso- tropic halos built from accretion of smaller galaxies;
the level of predicted anisotropy is consistent with results from a
few nearby ellipticals studied so far. In simulated BCG galaxies,
additional high-velocity components from disrupted or stripped
galaxies may be superposed on the central galaxy. Observationally, the
outer halos of the BCGs at the centers of the Fornax, Virgo, Coma and
Hydra I clusters are in fact harbouring substructure associated with
disrupted satellites. In all cases the data indicate that galaxy halos
and ICL are discrete components; the former do not blend continuosly
into the latter, and there is no evidence for a continuous increase of
the galaxy velocity dispersion to cluster values, but rather a colder
BCG halo is superposed with a hotter ICL component at radially
decreasing surface brightness ratio. The evidence for merging (in Coma)
and accretion/disruption indicates that the build-up of the BCG and ICL
is an on-going and long-lasting process. References: Gerhard et
al.\ (2007), Murante et al.\ (2007), de Lorenzi et al.\ (2009).

\subsubsection{The kinematics of intracluster planetary nebulae in the
  Hydra I cluster - G. Ventimiglia} The Hydra I cluster is a relaxed
cluster from its regular X-ray emission, in the Southern hemisphere at
50 Mpc distance. At this distance, even the emission line flux from
the brightest PN is only $8 \times 10^{-18}$ erg s$^{-1}$ cm$^{-2}$;
thus to detect these objects the sky noise must be substantially
reduced. This can be achieved with the ``multi slit imaging
spectroscopy'' technique (MSIS), a blind technique which combines the
use of a mask with parallel slits, a narrow band filter and a grism,
yielding spectra a few tens of \AA\ wide for all emission objects
which lie behind the slits.  With FORS2 and the VLT, in a 6.8
arcmin$^2$ squared field centred on NGC 3311, a total of 82 emission
line objects were identified: 56 are PN candidates and 26 background
galaxies. The $m_{5007}$ magnitudes are consistent with a PN
population at 50 Mpc distance, and the PN LOSVD shows three velocity
components: a main cluster component at the cluster's redshift and
expected $\sigma=600$ \kms, plus two discrete colder components at a
bluer (1800 \kms) and a redder (5000 \kms) velocity, providing
evidence for unmixed components in the NGC 3311 halo. References:
Gerhard et al.\ (2005), Arnaboldi et al.\ (2007), Ventimiglia et al.\
(2008).

\subsubsection{Planetary nebulae in NGC~1399: the kinematics of a cD
  halo - E. McNeil} A counter dispersed imaging spectroscopy study
with a mosaic of 5 pointings has been carried out, covering the bright
central parts and extended halo of the Fornax cD galaxy
NGC~1399. These observations deliver a sample of PN positions,
measured magnitudes and velocities, which can be used to construct a
2D velocity field and study the kinematics of the extended halo. In
this sample, 146 PNe associated with NGC~1399 are detected, 23 PNe
associated with NGC~1404, and 6 unassigned. From the projected phase
space diagram $v_{LOS}$ vs.\ radius, 12 PNe are identified which are
associated with a new low velocity component superposed onto NGC 1399.
The velocity dispersion profile at large radii in NGC 1399 is
consistent with a flat profile at 250 \kms; the rise to Fornax cluster
dispersion is not yet reached at these radii.  References: Arnaboldi
et al. (1994), Saglia et al. (2000), Mc Neil et al.\ (in preparation).

\section{Intracluster stellar populations - [Fe/H] and age 
distribution}\label{sec:pop}

\subsection{Planetary nebulae as tracers of stellar populations - R. Ciardullo} 
PNe can be powerful probes of the chemical history of stellar
populations. However, a full analysis of the $\alpha$-element
metallicity distribution function based on the nebular line ratios and
the detection of the temperature-sensitive [OIII]4363\AA\ line is not
yet feasible for intracluster studies. Still it is possible to search
for evidence of a metal-rich intracluster population with PN deep
spectroscopy, but such spectroscopic data must reach $\sim 20$ times
fainter than [OIII] $\lambda$5007 fluxes, and the next generation of
telescope may be crucial for this. PN counts (normalized to underlying
luminosity) do probe the stellar population, but we need a better
theoretical understanding of how these objects come to form. If the
blue straggler hypothesis is correct, then we may soon be able to use
PNe to measure/constrain the age of an old stellar
population. References: Feldmeier et al.\ (2004b), Ciardullo et al.\
(2005), Buzzoni et al.\ (2006).

\subsection{The intracluster red giant star population in the Virgo
  cluster - P. Durrell} 
\begin{figure}
\includegraphics[height=3in,width=3in]{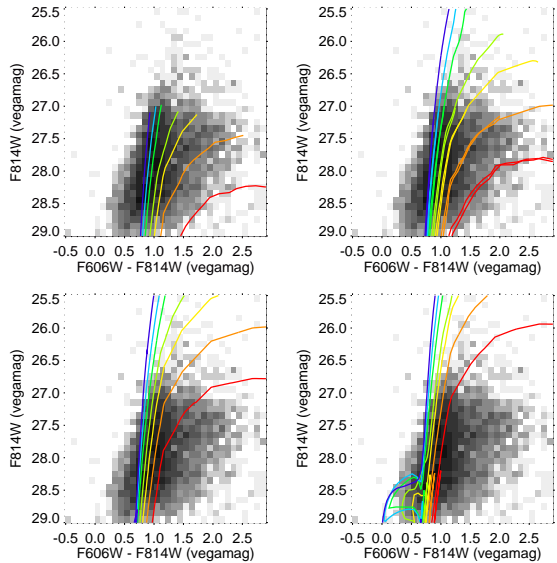}
\caption{Results from the VICs project. Observed colour magnitude
  diagram of the VICS Virgo field, with a subset of the Girardi et
  al.\ (2002) isochrones for the HST ACS filters superposed. In each
  panel, the isochrones represent stars with metallicities of Z =
  0.0001, 0.0004, 0.001, 0.0025, 0.004, 0.008, and 0.019, with redder
  isochrones corresponding to higher values of Z. Each panel displays
  a different age: log(age) = 10.1 (upper left; the contribution of
  the AGB was removed from these isochrones), 9.65 (upper right), 9.0
  (lower left), and 8.5 (lower right). From Williams et al.\
  (2007).}\label{fig:virgoRGB}
\end{figure}

Intracluster red giant branch (RGB) stars are
the most numerous visible component of ICL; with RGB stars, we can determine
the metallicities and constrain the ages of the stellar population(s)
(latter with AGB) and relate these to the galactic origin of diffuse
stellar light. The IC-RGB stars in Virgo can be studied only with very
deep imaging, e.g., HST, which implies a limited field of view.  The
study of the VICS field yielded over 5000 IC-RGB stars whose color
magnitude diagram shows a clear RGB populated by $\sim5300$ IC-RGB and
AGB stars, above a small background contamination, see
Fig.~\ref{fig:virgoRGB}. These intracluster RGB stars are mostly old
and metal-poor in the surveyed field, and they are not well-mixed even
on small scales. These results are consistent with stripping of stars
from a wide variety of galaxies, from dwarfs to the outer regions of
ellipticals and spirals. References: Durrell et al.\ (2002), Williams
et al.\ (2007).

\subsection{Intracluster globular clusters and ultra-compact dwarfs -
  M. Hilker} Galaxy clusters are the places of highest galaxy density
in the local universe, their cores are dominated by giant ellipticals
with an extended surface brightness envelope (cD halo).  Globular
clusters are excellent tracers of diffuse stellar populations due to
their abundant numbers and brightness.  The central cluster galaxies
possess extremely rich globular cluster systems ($\sim ~10.000$ GCs)
with an equal amount of red (metal-rich) and blue (metal-poor) GCs.
The predominant GC population in intra-cluster regions are the blue
(metal-poor) GCs which can be found out to large cluster-centric
radii.  Also very massive star clusters/ultra-compact dwarf galaxies
(UCDs, $10^6-10^8 M_\odot$) are found in the intra-cluster space. Some of
these might have had their origin as nuclei of now disrupted dwarf
galaxies. References: Richtler et al.\ (2004), Schuberth et
al.\ (2009), Misgeld et al.\ (2009).

\subsection{Intergalactic globular clusters - M. West} 
Numerical simulations suggest large populations of intergalactic
globulars could exist in rich galaxy clusters. There is evidence of
wide-spread galaxy stripping and destruction. The Jeans mass at
recombination was $\sim10^5 - 10^6$ solar masses, and therefore globular
clusters could form anywhere the mass density was high enough.  The
Abell cluster 1185 provides clear evidence for IC GCs found near the
peak of the Xray emission which does not coincide with the center of
the cluster galaxy number density distribution. The deep ACS image at
this position shows an excess of point-like sources with respect
to the control fields.  A large population of IGCs ($\sim 50,000$ GCs)
is observed in Coma, while the Virgo cluster has a small
population. References: Jordan et al.\ (2009), Takamiya et al.\
(2009).

\subsection{Stellar kinematics and line strength indices in BCG halos - 
L. Coccato}
The bright galaxies at the centers of galaxy clusters often have
extended halos whose presence is considered to be the result of the
co-evolution of the central galaxy and its cluster
environment. Information on the time of formation of the different
components comes from the study of their stellar populations, which
may provide both age and metallicity profiles.  Long slit absorption
line spectroscopy of the two BCG galaxies in the Coma cluster,
NGC~4889 and NGC~4874, was performed at the Subaru 8 m telescope, with
the FOCAS spectrograph.  The integrated light spectra had adequate S/N
so that absorption line indices for NGC 4889 could be determined out
to 65 kpc. The results show that within $1.2 R_e$ the stars of
NGC~4889 must have formed over a very short time scale ($< 1$ Gyr),
with little subsequent merging, as shown by the large
$\alpha-$enhancement and the measured steep metallicity gradient. The
stars in the halo were formed on a longer time scale, as shown by the
lower $\alpha$/Fe values. This is compatible with formation in smaller
systems which were subsequently accreted onto the halo of the BCG.
References: Coccato et al.\ (in preparation).

\section{Cosmological simulations of cluster and group 
formation/Origin of diffuse light}\label{sec:simulc}

\subsection{Cluster and group formation in $\Lambda$CDM - S. White} 
Starting from cosmological $\Lambda$CDM initial conditions, numerical
simulations of the formation of large scale structure can now predict
and reproduce the average mass profiles and properties of clusters and
groups in the mass range $10^{12}\msun$-$10^{15}\msun$.  Table 1 shows
some parameters for the Millenium-II simulation. Together with
semi-analytic models the simulations can be used to study the stellar
mass function of galaxies in groups and clusters down to $10^7\msun$,
which is found to have similar shape as in the field except for the
BCGs.  Assuming the ICL is due to tidally stripped and disrupted
galaxies, based on galaxy orbits and sizes and stripping of the DM
halo, the simulations can be used to predict the fraction of cluster
light in the ICL. Considerable scatter in the fractions of ICL
relative to ICL + BCG (within $\sim 40\pm30\%$) and relative to the
total cluster light (within $\sim 15\pm15\%$) is found, due to
variations in assembly/stripping history. References: Springel, Frenk
and White (2006), Hilbert and White (2009), Guo et al.\ (2009).
\begin{table}[h]\def~{\hphantom{0}}
  \begin{center}
  \caption{The Millennium-II simulation: particle number and mass,
    size of simulated volume, resolution (softening), and fraction
    of particles in lumps at z=0.}
  \label{tab:MS}
  \begin{tabular}{lcccc}\hline
     $N_{part}$ & $m_{part}$ & L & $\epsilon$ & F$_{halo}(z=0)$
    \\ & $h^{-1}M_\odot$ & $h^{-1} Mpc$ & $h^{-1}$ kpc &
    \\ \hline
2160$^3$ & $6.9\times 10^6$ &
    100 & 1 & 0.60\\\hline
  \end{tabular}
 \end{center}
\end{table}

\subsection{Galaxy evolution in clusters and groups - L. Mayer} 
The physical mechanisms for galaxy transformation at work in cluster
environments can be grouped into two categories: tidal interactions,
and ICM-galaxy interactions. The latter can take place in several
flavours: ram pressure stripping of cold ISM in disks; triggered star
formation and strangulation; stripping of gaseous halos around
galaxies. These processes help to build up the morphology density
relation, i.e., that early-type galaxies (E,S0,dE,dS0) dominate in
cluster cores, and explain the high number density of early-type dwarf
galaxies at the centers of nearby groups and clusters, the HI
deficiency of galaxies in cluster centers, and the steepening of the
faint end of the luminosity function relative to the field. These
transformations were studied with simulations replacing dark halos in
dark matter only simulations at the time of infall with
multi-component late type galaxy models.  Cosmological simulations of
groups show that mergers drive morphology towards elliptical systems,
and that at large $R>5R_e$ a diffuse stellar halo or intragroup light
component is formed which may comprise $\sim 20\%$ of the group light.
References: Mastropietro et al. (2005), Mayer et al. (2006).

\subsection{Galaxy properties in clusters: dependence on the environment 
- H. Muriel} 
This work investigated the dependence of several galaxy properties on
the environment and cluster identification techniques. Clusters of
galaxies were selected from two catalogues based on the SDSS: the
ROSAT-SDSS Galaxy Cluster Survey, and the MaxBCG Catalogue.  Galaxies
in X-ray and MaxBCG selected clusters show similar size-luminosity
relations.  The Faber-Jackson relation for early-type galaxies in
clusters is also the same for X-ray selected and MaxBCG
clusters. BCGs, non-BCG-early type galaxies in clusters and field
early-type galaxies show different size-luminosity relations and have
different dynamical properties.  Using several criteria to classify
galaxies into morphological types, the well known morphological
segregation can be reproduced. These results can be related to the
different processes that affect the evolution of galaxies in various
environments. References: Coenda and Muriel (2009).

\begin{figure}
\includegraphics[height=1.7in,width=5.1in]{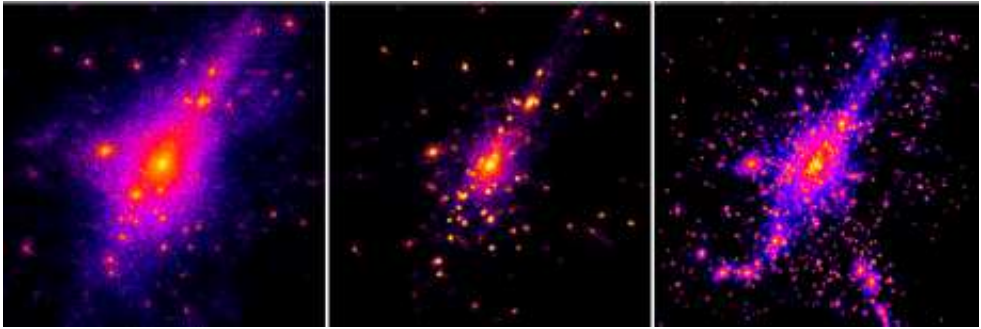}
\caption{The distribution of the dark matter (DM, left), and of the
  stars (center), for a simulated cluster, and the distribution of
  stars in a high-resolution re-simulation of the same cluster
  (right), all at redshift $z=0$. The frames are $6 h^{-1}$ Mpc on a
  side, corresponding to $\approx 2 R_{\rm vir}$. They show density
  maps generated with different logarithmic colour scales for DM and
  stars. The galaxy and diffuse components in each panel can clearly
  be seen. From Murante et al.\ (2007).  }
  \label{fig:simcluster}
\end{figure}

\subsection{Diffuse stellar component in cosmological simulations - K. Dolag} 
Cosmological hydro-dynamical simulations of zoomed-in clusters have
been performed with up to $4\times 10^6$ particles in $R_{vir}$,
radiative cooling/heating by UV background (quasars), a sub-grid
two-phase model for star-formation, thermal and kinetic feedback by
supernovae; and evolving stellar populations (SNIa/SNII).  In these
simulations it is possible to separate galaxies, the BCG, and the ICL,
using substructure finding algorithms and the full phase-space
information of the stellar particles.  The ICL surface brightness
profile is found to be steeper than the average galaxy light from
cluster galaxies averaged in annuli.  From the analysis of the merger
trees in these cluster simulations, no preferred formation time for
the ICL is found, and because it is cumulative, the largest
contribution is released at $z<1$. Most of the ICL comes from merging
processes, either with the BCG or prior to infall into the cluster;
other contributions are significant only outside $0.5
R_{vir}$. Mergers liberate up to 30\% of stars in both
galaxies. Related with the finite resolution of the simulations, a
sizable contribution comes from dissoved galaxies, but whether this is
a robust estimate cannot be firmly established at this
stage. References: Murante et al.\ (2004, 2007), Dolag et al.\ (2009).

\subsection{Extended ionized and molecular gas emission in galaxy
  clusters - R. Oonk } New, deep integral-field spectroscopy and
imaging of the extended molecular and ionized gas distributions within
the central regions of several galaxy clusters is presented, obtained
with the VLT and HST. These observations show the existence of gas
surrounding the BCG, extending at least up to 20 kpc from the
nucleus. The H2 to HII line ratios are very high and are different
from typical AGN and starburst ratios. To date no single source of
excitation has successfully explained all line ratios over the
entirety of the observed gas distribution. Various line diagnostics
are used to constrain the properties of the observed gas and discuss
its origin and fate. The question is open whether this is a birthplace
of some of the stars in the ICL. References: van Weeren et al. (2009).

\section{Summary of the discussion}

The discussion at the end of the workshop was introduced by
M.~Arnaboldi with a quote from Uson et al.\ (1991): ''Whether this
diffuse light is called the cD envelope or diffuse intergalactic light
is a matter of semantics; it is a diffuse component which is
distributed with elliptical symmetry about the center of the cluster
potential''. A lively discussion then developed about the progress
achieved since the time of this paper, with the main contributors
being M.~Arnaboldi, R.~Ciardullo, P.~Durrell, J.~Feldmeier, C.~Mihos,
S.~White, and A.~Zabludoff. In the following, we list the main topics
of discussion, and the specific conclusions and open issues that
emerged for each of them. Because this list reflects only our
understanding of what was being said, we apologize in advance for any
errors or omissions.

\parskip=5pt
\par\vskip5pt\noindent{\scshape 
Envelope of brightest cluster galaxy versus ICL:} 

\bullitem Sometimes the ICL is not aligned with BCG isophotes, so from
this fact alone one would conclude that cD halo and ICL are different
components.

\bullitem In some clusters like Coma and Abell 1185 an extended ICL
component embeds more than one BCG. In Abell 545, a bright diffuse
light component is seen with several galaxies embedded, but no cD.
The ICL must therefore be a component distinct from a galaxy halo.

\bullitem There is kinematic evidence from PN and GC velocity
distributions in the centers of nearby galaxy clusters that ICL and
BCG halos are distinct physical components.

\bullitem There is a clear need for more spectroscopic data, to
disentangle the contribution of halo and cluster component.

\par\vskip7pt\par\noindent{\scshape How to measure the fraction of ICL?}

\bullitem The fraction of ICL is an observationally ill-defined
quantity, either depending on arbitrary surface brightness thresholds
in photometric studies, or lacking the full phase-space information
to ascertain whether a star is bound to the central galaxy or not. A
comparison with simulations is therefore difficult.

\bullitem Thus it would be useful if simulations were analysed to
produce surface brightness maps. This would facilitate the comparison
with wide field photometry measurements. One would then adopt a
surface brightness threshold and make comparisons between the observed
and simulated ICL plus BCG envelope fractions.

\bullitem A useful measurement to quantify the dynamical status of the
ICL would be the fraction of homogenous versus irregular light or,
more generally, the power spectrum over spatial scales.

\par\vskip7pt\par\noindent{\scshape To what redshift can we go with
ICL studies?}

\bullitem Since much of the merging of massive galaxies occurs from
z=1 to z=0, it would be interesting to reach z=0.3 with studies of ICL
morphology, photometry and kinematics.  Redshifts higher than z=0.3
may be out of reach because of the strong surface brightness dimming
effect scaling $\propto(1+z)^{-4}$ with redshift $z$.

\bullitem The deep data sets obtained in weak lensing surveys may be
useful for the determination of ICL morphology and photometry in
intermediate redshift clusters.

\par\vskip7pt\par\noindent{\scshape Metallicity and colour as 
observational constraints on how ICL is made:}

\bullitem In the VICS field in Virgo the intracluster stars
are mostly old and mostly metal poor with a large metallicity
spread; this is consistent with stripping of stars from a
wide variety of galaxies, including the halos (but not the
inner regions) of ellipticals. 

\bullitem Most of the ICL in moderate redshift clusters, that is the
more nearly homogeneous part, has the same color as the early-type
galaxies in the cluster core.

\bullitem Blue colors are measured most often for elongated
stream-like features.

\bullitem It would be very important to use cosmological simulations
to predict the metallicity distribution and colours of ICL stars, and
to compare this with present data.

\par\vskip7pt\par\noindent{\scshape Results from simulations on the origin
  of ICL:}

\bullitem In cosmological simulations of cluster formation, most of
the ICl comes from the halos of evolved galaxies. Does this need to be
reconciled with the results from higher-resolution simulations of
galaxy evolution in dense environments?

\bullitem The importance of groups in generating ICL must be
emphasized, through creating loosely bound stars in group interactions
(``pre-processing'') which are later unbound by the gravitational
potential of the cluster into which they fall.

\bullitem Groups are generally important because of their
environmental effects on galaxy evolution, and thus indirectly, the
ICL.

\par\vskip7pt\par\noindent{\scshape The high-mass end of the galaxy
luminosity function:}

\bullitem Is there still a problem caused by the evolution of the high
mass end of the galaxy mass function from redshift z=1 to z=0?

\bullitem With semi-analytical models one can adjust the parameters
for the AGN feedback, and recover agreement with observations. In the
case of hydro-dynamical simulations, there is still a problem, and the
dispersion of stars in the cluster volume as ICL would help.

\par\vskip7pt\par\noindent{\scshape Future prospects:}

\bullitem Currently we see a lot of new data on intracluster globular
clusters from the Virgo/Coma legacy surveys. It will require lots of
spectroscopic time to do the kinematic follow-up observations but the
result will definitely be worth it.

\bullitem Resolved stellar populations in the Virgo and Fornax cluster
ICL are obvious targets for the E-ELT, JWST, because crowding is not a
problem at these low surface brightness levels.

\par\vskip7pt\par\noindent{\scshape Concluding remark:}
Jorge Melnick summarized the meeting by stating that he was very
pleased to see the subject of ICL to be active and strong, with
extensive developments in deep surface brightness measurements,
kinematics, stellar population studies, and comparisons with
cosmological simulations. A very interesting meeting, much new
science, and hope for more in Beijing!

\section{Poster Papers}
\begin{itemize}
\item ABS N. 123: Eduardo Cypriano et al. - Shrinking of cluster
ellipticals: a tidal stripping explanation and implications for the
intracluster light
\item ABS N. 210: Marcelo Bryrro Ribeiro - Differential density
statistics of the galaxy distribution and the luminosity function
\item ABS N.450: Margarita Rosado et al. - Diffuse light in the
Seyfert's Sextet
\item ABS N.855: Tiberio Borges Vale et al. - Environmental
effects on the structure of galaxy discs
\item ABS N.992: Cristina Furlanetto et al. - Detection of
gravitational arcs in galaxy clusters
\item ABS N.1877: Walter Augusto Santos et al. - Photometric
redshifts for SDSS galaxies using locally weighted regression.
\item ABS N.1951: Julio Saucedo et al. - A study of the galaxy
population in the region of A1781
\item ABS N. 1962: Yasuhiro Hashimoto et al. - Multi-wavelength
study of cluster morphology and its implications on the scaling
relations, mass estimate, large scale structure, and evolution of
galaxies.
\item ABS N. 2076: Steven Michael Crawford et al. - The evolution of
cluster galaxies and the diffuse intracluster light
\item ABS N. 2341: Simon Nicholas Kemp et al. - From cDs to diffuse
structures: faint light in galaxies using Schmidt and CCD data
\item ABS N. 2728: Mangala Sharma et al. - Tracing galaxy group
dynamical histories through diffuse intergalactic light.
\item ABS N. 2947: Sadanori Okamura et al. - Observation of
intracluster diffuse light in the Coma cluster
\item ABS N. 2990: Roderik Overzier - Examining the Spiderweb:
forming a BCG and its intracluster light at z=2
\item ABS N. 3002: Nieves Castro-Rodriguez et al. - Intracluster
light in the Virgo cluster: large scale distribution.
\end{itemize}

\begin{acknowledgments}
  We would like to acknowledge the support from the IAU, and the hard
  work and enthusiasm of the participants, which made this IAU Joint
  Discussion \#2 on Diffuse Light in Galaxy Clusters both possible and
  stimulating.
\end{acknowledgments}

\end{document}